\documentclass[prl,a4paper,twocolumn,showpacs,amssymb]{revtex4}
\usepackage{graphicx}

\begin {document} 

\title{The superfluid Reynolds number and the transition from potential flow to turbulence in superfluid $^4$He at mK temperatures}

\author{W. Schoepe}

\affiliation{Fakult\"at f\"ur Physik, Universit\"at Regensburg, D-93040 Regensburg, Germany}

\begin{abstract}
This comment is on Phys.Rev.Lett.\,144, 155302 (2015) by M.T. Reeves, T.P. Billam, B.P. Anderson, and A.S. Bradley ``Identifying a superfluid Reynolds number via dynamical similarity'' where a new superfluid Reynolds number is introduced. This definition is shown to be useful in the data analysis of the finite lifetime of turbulence observed with an oscillating sphere in superfluid helium at mK temperatures in a small velocity interval $\Delta v =(v-v_c)$ just above the critical velocity $v_c$. The very rapid increase of the lifetime with increasing superfluid Reynolds number is compared with the ``supertransient'' turbulence observed in classical pipe flow. 

\end{abstract}


\maketitle 
\textbf{1.\,Introduction.}
In classical hydrodynamics the concept of the Reynolds number $v D/\nu$  (where $v$ is the flow velocity, $D$ is the size of the object in the velocity field, and $\nu$ is the kinematic viscosity of the liquid) is very useful because it reflects the similarity properties of the Navier-Stokes equation. For example, the drag coefficient of a sphere moving through the liquid is a universal function of the Reynolds number only, irrespective of individual values of $v, D,$ and $\nu$.\cite{LL}

In a superfluid the situation is more complex because there are three different velocities, namely the velocity of the superfluid component $v_s$, the velocity of the normal component $v_n$, and the motion of the quantized vortices with respect to $v_s$ and $v_n$, the latter one was shown to be an intrinsic property being independent of the flow velocities, but only dependent on the interaction between both components. In superfluid $^3$He those parameters lead to three different Reynolds numbers.\cite{Finne} 

In superfluid $^4$He below 0.5 K the normal component is almost absent, only a dilute gas of ballistically propagating phonons is left. Therefore, a hydrodynamic velocity $v_n$ does no longer exist. In addition, it was shown in Ref.[2] that at these low temperatures the motion of the vortices in $^4$He is undamped and, hence, they are subject to turbulent motion. As a result, only one ``superfluid Reynolds number'' $Re_s$ remains to be defined, namely by replacing the kinematic viscosity $\nu$ by the circulation quantum of the vortices $\kappa = h/m$ (where $h$ is Planck's constant and $m$ is the mass of a helium atom) because $\kappa$ has the same dimension m$^2$/s as the viscosity:\,\,$Re_s = v D/\kappa$.\cite{Res} It describes the ratio of an overall circulation to that of one vortex. In fact, in several theoretical investigations turbulence has been found for $Re_s\gg$ 1.\cite{Res>>1}

In the new work by Reeves et al.\,\cite{Reeves} the two-dimensional Gross-Pitaevskii equation is investigated numerically in the vicinity of the critical velocity $v_c$ for the onset of quantum turbulence. The central result of that work is the observation of dynamic similarity in the wake of a cylindrical object and its breakdown due to vortex shedding at a superfluid Reynolds number defined as
\begin{equation}
Re_s \equiv \frac{(v-v_c)D}{\kappa}.
\end{equation}
(The authors define here $\kappa = \hbar/m$; in the following we continue to use $h/m$.) Turbulence is observed in the wake at $Re_s \approx$ 0.1, irrespective of cylinder size.\\ 

\textbf{2.\,The critical velocity \boldmath{$v_c$}.}
In order to prove that (1) is a useful definition for the onset of turbulence in superfluid helium, at first a determination of the critical velocity $v_c$ is necessary. It is known that $v_c$ depends on the experimental setup.\\ 

Firstly, in experiments with steady flow the Feynman critical velocity $v_c \sim \kappa /D$ is expected for $v\,D/\kappa \sim$ 1. Inserting this into (1) gives

\begin{equation}
Re_s \sim \left(v - \frac{\kappa}{D}\right)\,\frac{D}{\kappa} \sim  \frac{v\,D}{\kappa} -1.
\end{equation}         
Only the zero point is shifted between both superfluid Reynolds numbers. There is no difference when $Re_s \gg$1 or $v\gg v_c$.

Secondly, in practice experiments with oscillatory flow are more easily performed, are more reliable, and in general yield sharp critical velocities. In the experiments with oscillating spheres \cite{spheres}, wires \cite{wires}, and tuning forks \cite{forks} at various frequencies $\omega/2\pi$ ranging from ca.\,100 Hz up to 30 kHz
a critical velocity has been measured that scales as $v_c \sim \sqrt{\kappa \omega}$ (neglecting at the moment numerical prefactors of order 1). Considering the oscillation amplitude $v/\omega$ as the relevant length scale $D$ in $v\,D/\kappa \sim$ 1 we obtain the experimental scaling. A rigorous derivation, however, is not available because there is still no theory of the transition to turbulence in oscillatory superflows.\\

\textbf{3.\,Transition to turbulence around an oscillating sphere at \boldmath{$v_c$}.}
Equation (1) is obviously relevant when the velocity amplitude is close to $v_c$, in particular when $\Delta v/v_c \ll$ 1. This is the case in our experiments with an oscillating sphere (radius $R$ = 0.12 mm, frequencies 119 Hz and 160 Hz, $v_c \sim$ 20 mm/s) at helium temperatures below 0.5 K.\cite{wir1}\, In a small interval of $\Delta v =(v-v_c)$ of few percent above $v_c$ we
observe that the flow pattern is unstable, switching intermittently between turbulent phases and potential flow. These patterns are easily identified because the drag force on the sphere is much larger in the turbulent regime than during potential flow. Recording time series at constant temperature and driving force we analyze the distribution of the lifetimes of both phases. The turbulent phases have an exponential distribution $\exp(-t/\tau)$ and the mean lifetimes $\tau$ increase very fast as we increase the velocity amplitude, namely as
\begin{equation}
\tau = \tau_0\,\exp\,(\Delta v/v_1)^2,
\end{equation}  
\begin{figure}[t]
\centerline{\includegraphics[width=1.3\columnwidth,clip=true]{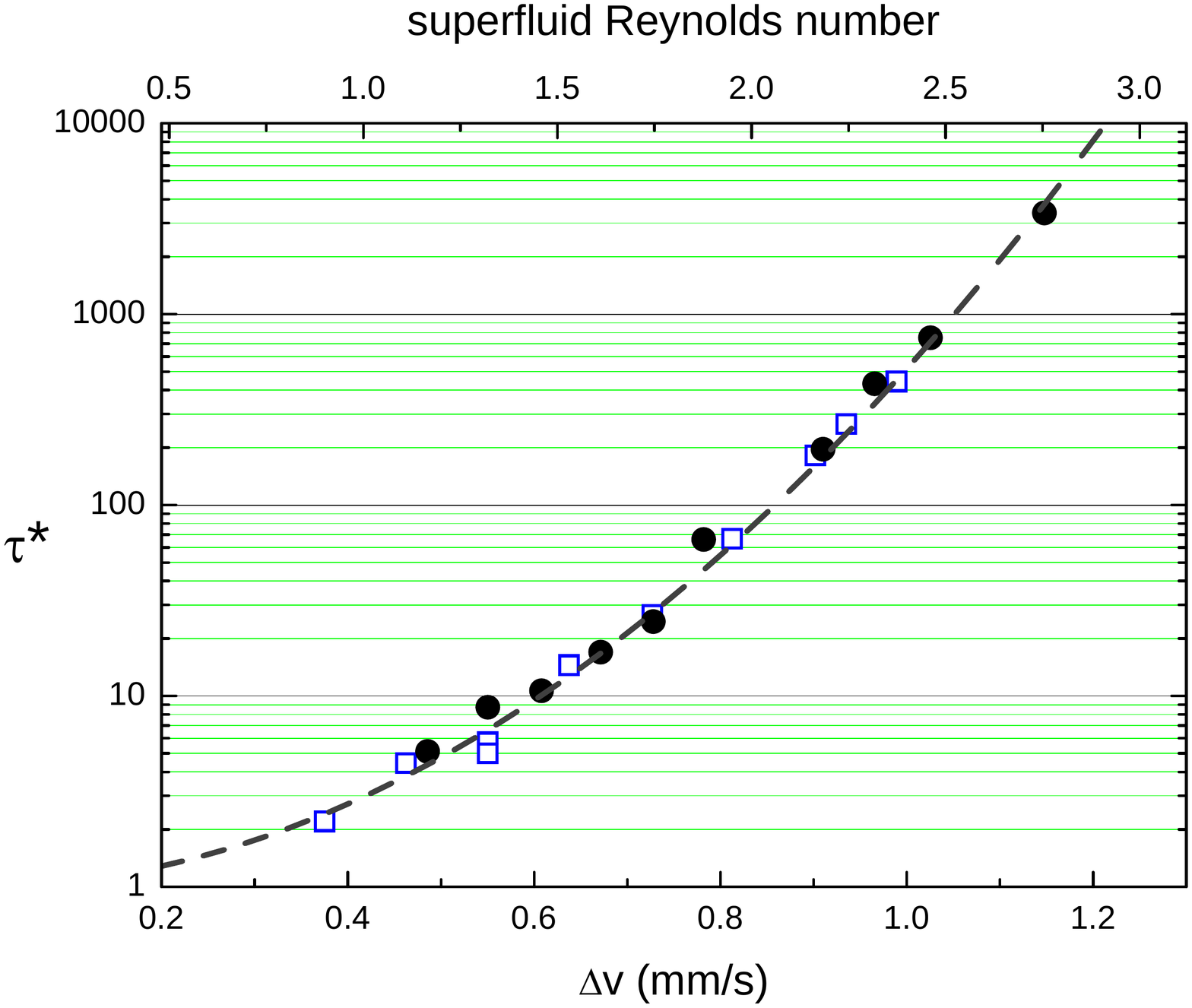}}
\caption{(Color online) The normalized lifetimes $\tau ^* = \tau /\tau_0$ as a function of $\Delta v = v - v_c$ for the 119 Hz oscillator at 301 mK (blue squares) and the 160 Hz oscillator at 30 mK with 0.05\% $^3$He (black dots). Note the rapid increase of $\tau ^*$ by 3 orders of magnitude over the small velocity interval of ca. 0.7 mm/s\,. The frequency, the temperature, and the $^3$He concentration have no effect on the data. The dashed curve is calculated from Eq.(3).}
\label{fig:1}       
\end{figure}\\
where $v_1$ = 0.48 $\kappa /R$, and $\tau_0$ = 0.5 s at 119 Hz and 0.25 s at 160 Hz \cite{wir1}. 
The velocity $v_1$ is of the same order as the Feynman critical velocity. The numerical prefactor of 0.48 is due to various numerical factors which can safely be considered to be independent of the size of the sphere. Only the prefactor in $v_c = 2.8 \,\sqrt{\kappa \omega}$ \,has not been proven to be independent because the size of the sphere was not varied. However, with oscillating wires  \cite{Yano} the prefactor of $v_c$ is 2.1, and this can be understood simply when comparing the maximum velocity at the equator of the sphere that is 1.5 times the fluid velocity far away with the maximum velocity at the circumference of a cylinder that is twice the velocity far away, therefore 2/1.5 = 1.33 and 2.8/2.1 = 1.33. Hence, we expect the prefactor to have no significant dependence on the size of the sphere, if any.

In Fig.1 we plot the normalized lifetime $\tau^* \equiv \tau /\tau_0$ vs. $\Delta v$. The salient feature is that $\tau^*$ is independent of the oscillation frequency, of the temperature, and is not affected by $^3$He impurities.
The only frequency dependence is in $\tau_0$. The exponential dependence of $\tau $ on $(\Delta v / v_1)^2$ may be described by Rice's formula \cite{Rice,Tikh} for Gaussian fluctuations crossing a given level. It is applied here to fluctuations of $v$ crossing $v_c$ where turbulence breaks down. The average number per unit time of crossings of a given level $C$ below a mean at zero with negative slope is proportional to $\exp(- C^2/2\sigma ^2)$ with $\sigma ^2$ being the variance of the fluctuations. Postulating that the lifetime of a turbulent phase is ending at the first crossing of $v$ of level $v_c$ , we obtain (3) \cite{PRL}. $\tau _0 $ is interpreted to be twice the average time between two crossings of the level $v_c$ \cite{Rice}, i.e., for $\tau _0$ = 0.5 s or 0.25 s there are 4 or 8 crossings per second which means that the velocity of the superfluid around the oscillating sphere fluctuates at $v_c$ with an average frequency of 2 Hz or 4 Hz, respectively. This could not be detected with our electronics. So far, we have no explanation for this result.\\ 

\textbf{4.\,Supertransient turbulence in $^4$He.}
Turbulence having a finite lifetime before breakdown to laminar flow is called ``transient'' turbulence, and in case of lifetimes growing faster than exponentially with Reynolds number the term ``supertransient'' is used.\cite{Tel} An example of supertransients was observed in classical pipe flow where the lifetimes of turbulent ``puffs'' have a double-exponential dependence on the Reynolds number.\cite{Hof}

In our case we can replace $\Delta v$ by $Re_s$ in (3) and set $D$ = 2$R$ to obtain
\begin{equation}
\tau^* = \,\exp\,(c\, Re_s)^2,
\end{equation}\\ 
with c = 1.04. In Fig.1 the dependence of $\tau^*$ on $Re_s$ is shown at the top axis (note that $Re_s$ = 0 corresponds to $\Delta v$ = 0 and $\tau^*$ = 1). It is obvious from Fig.1 that our values of $Re_s$ are all below 3, corresponding to maximum mean lifetimes $\tau$ of about 15 min and to $\Delta v/v_c \leq$ 0.04. An extension towards larger velocities would require measuring times lasting much longer than our longest time series of 36 hours. Therefore, the validity of (4) cannot safely be extended to larger $Re_s$. Nevertheless, we can state that 
the fast increase of $\tau^*$ with $Re_s$ is another example of supertransient behavior, here for the first time in a pure superfluid.\\    

\textbf{5.\,Conclusion.}
The new superfluid Reynolds number is indeed a useful quantity for the investigation of the transition to turbulence in oscillatory flow of superfluid helium just above the critical velocity $v_c$. The turbulent lifetimes grow very fast but without diverging at some particular superfluid Reynolds number. When the lifetimes begin to exceed the time of experimental observation, turbulence only $\it{appears}$ to be stable, but actually it is supertransient chaos.\\

\textbf{Acknowledgement.}
It is a pleasure to acknowledge fruitful discussions with Risto H\"anninen (Aalto University, Finland), with Michael Niemetz (OTH Regensburg, Germany), and especially with Grigori Volovik (Aalto University, Finland and Landau Institute for Theoretical Physics RAS, Russia).

\bigskip
e-mail:wilfried.schoepe@ur.de



\begin{thebibliography}{99}
\bibitem{wir1}
W. Schoepe, J. Low Temp. Phys. \textbf{173}, 170 (2013);
arXiv:1309.1956 [cond-mat.other]. In that work we found $v_1$ = 0.41\,$\kappa/R$ but we also mentioned that, strictly speaking, a 15 \% correction is needed when inferring the quadratic drag force from the driving force. Taking this into account the prefactor of $v_1$ changes from 0.41 to 0.48.
\end{thebibliography}
\end{document}